\documentclass[nofootinbib, reprint, preprintnumbers, amsmath, amssymb, amsfonts, aps, pra, superscriptaddress, floatfix]{revtex4-2}
\usepackage[utf8]{inputenc}
\usepackage{mathtools}
\usepackage{graphicx}
\usepackage{dcolumn}
\usepackage{bm}
\usepackage{physics}
\usepackage{float}
\usepackage{placeins}
\usepackage{color}
\usepackage{braket}
\usepackage[normalem]{ulem}
\usepackage{enumerate}
\usepackage{enumitem}
\usepackage[colorlinks=true,linkcolor=blue,citecolor=blue,urlcolor =blue]{hyperref}
\usepackage{mathrsfs}
\usepackage{bbm}
\usepackage{siunitx}
\usepackage{graphicx,color}
\usepackage{amssymb}
\usepackage{amsmath}
\usepackage{placeins} 
\usepackage{subcaption}
\usepackage{bm}
\usepackage{comment}
\usepackage{lipsum}

\begin{document}
\title{
Proposal for Estimating the Energy Gap of the Transverse-Field Ising Hamiltonian Using a D-Wave Quantum Annealer
}

\author{Kota Yamada}
\affiliation{Department of Electrical, Electronic, and Communication Engineering, Faculty of Science and Engineering, Chuo university, 1-13-27, Kasuga, Bunkyo-ku, Tokyo 112-8551, Japan}%




  

\author{Yuichiro Matsuzaki}
\email{ymatsuzaki872@g.chuo-u.ac.jp}
\affiliation{Department of Electrical, Electronic, and Communication Engineering, Faculty of Science and Engineering, Chuo university, 1-13-27, Kasuga, Bunkyo-ku, Tokyo 112-8551, Japan}%

\date{\today}


\begin{abstract}
The transverse-field Ising model is a fundamental quantum spin system that captures the competition between quantum fluctuations and interactions, playing a central role in studies of quantum phase transitions and non-equilibrium dynamics. However, classical computations of ground and excited states in large-scale or high-dimensional systems are severely limited by the exponential growth of the Hilbert space. Here, we propose a novel approach using a D-Wave quantum annealer, where a triangular-wave oscillating magnetic field is applied to induce Rabi oscillations, allowing the estimation of energy gaps between the ground and excited states. Unlike conventional quantum annealing methods limited to ground-state searches, this approach can directly access excited-state information. 
It is potentially applicable to larger systems, 
providing a new avenue for quantum-device-based simulation. The validity of the method is demonstrated through numerical simulations of relatively small systems.
\end{abstract}

\maketitle
\section{Introduction}

The transverse-field Ising model is one of the most fundamental quantum Hamiltonians. It is well known that this model can describe quantum phase transitions and nonequilibrium dynamics through the competition between quantum fluctuations and spin-spin interactions~\cite{Sachdev1999,Dutta2015,Suzuki2013,Dziarmaga2010,Polkovnikov2011}. Owing to these properties, it has been widely employed in the study of phase transitions and critical phenomena in one- and two-dimensional systems, and has become a standard model in condensed matter physics and statistical mechanics. 

In recent years, the search for the ground state of the transverse-field Ising model has also been applied to combinatorial optimization problems, making it an important framework in the development of quantum computing technologies and quantum algorithms\cite{Finnila1994,Kadowaki1998,Farhi2000,Das2008,Johnson2011,Lucas2014,Farhi2014,Albash2018}.
To understand the properties of the ground and excited states of the transverse-field Ising model, information about the eigenvalues and eigenstates of the Hamiltonian is indispensable. Conventionally, numerical methods such as exact diagonalization, the Lanczos method, and the density matrix renormalization group (DMRG) have been widely used~\cite{White1992}\cite{Schollwock2005}. However, these methods suffer from a fundamental limitation: the dimension of the Hilbert space increases exponentially with system size. For an Ising model consisting of $L$ spins, the Hilbert-space dimension is given by $2^L$. Once $L$ exceeds several tens of spins, the memory requirements and computational cost increase dramatically, making analysis using classical computers practically infeasible. Consequently, obtaining exact information about the ground and excited states of large-scale or high-dimensional systems remains a challenging problem.
To alleviate these limitations, approximate methods such as tensor-network approaches and quantum Monte Carlo techniques have been developed\cite{Orus2014}\cite{Sandvik2010}. Tensor-network methods are particularly effective for one-dimensional systems or systems with local interactions, as they can efficiently represent the entanglement structure of many-body quantum states. On the other hand, quantum Monte Carlo methods are powerful tools for investigating finite-temperature systems and thermodynamic properties of many-body systems. Nevertheless, these approaches also face significant difficulties when applied to higher-dimensional systems, strongly correlated systems, or systems suffering from severe sign problems. Therefore, no universally applicable classical method currently exists for the large-scale analysis of the transverse-field Ising model.

\textcolor{black}{
In principle, the energy gap could be measured using quantum annealing devices. Although experimental methods have been proposed for measuring the energy gap during quantum annealing, they require the preparation of a superposition of the ground and excited states \cite{Russo2021,Matsuzaki2021}. In the current implementation of D-Wave devices, however, preparing such a superposition is not possible. There is another method for measuring the energy gap during quantum annealing \cite{Mori2024}. However, this method requires oscillating the coupling strength between qubits, which makes experimental demonstration challenging. Therefore, a method capable of measuring the energy gap using currently available D-Wave devices is highly desirable.}

Here,
we propose a new approach for obtaining information about excited states of the transverse-field Ising model without relying on classical computational resources. Specifically, we employ a D-Wave quantum annealer\cite{Johnson2011,Dickson2013,Boixo2013,Boixo2014,Lanting2014} and apply a triangular-wave oscillating magnetic field, which acts as a pseudo drive, to a prepared ground state. By inducing Rabi oscillations, the energy gap between the ground state and excited states can be estimated. Although current D-Wave hardware cannot directly apply microwave fields, it can implement the triangular-wave magnetic-field modulation described later, which forms the basis of our approach.

Unlike conventional quantum annealing methods that are primarily designed for ground-state searches, our method enables the direct extraction of physical information related to excited states. In particular, by controlling the frequency and waveform of the applied oscillating magnetic field, it becomes possible to investigate the dynamical response of the system in detail. A major advantage of this method is that, in principle, it can estimate energy gaps even for large-scale Hamiltonians that are difficult to diagonalize using classical computers. This capability opens a pathway to accessing physical properties of systems that are beyond the reach of conventional numerical simulations, thereby providing a new framework for quantum-device-based quantum simulation. In this work, we verify the validity of the proposed method by numerically simulating the dynamics of a D-Wave quantum annealer for systems with a relatively small number of qubits using classical computation.

\section{Rabi Oscillations}

We begin with a brief review of Rabi oscillations. Rabi oscillations refer to the phenomenon in which the state of a two-level quantum system oscillates under the influence of an external driving field. In the presence of 
external oscillating field, the quantum state evolves in time and repeatedly transitions between specific states.
Even in a quantum system with many energy levels, Rabi oscillations can be induced by applying an oscillating magnetic field that is resonant with a selected pair of levels. We consider such a situation and introduce the following Hamiltonian:
\begin{align}
\hat{H} &= \hat{H}_0 + \Lambda \cos{\omega t}\hat{A} \\
\hat{H}_0 &= \sum_{n=0}^{N-1} E_n \ket{E_n}\bra{E_n}.
\end{align}
Here, $\hat{H}_0$ denotes the Hamiltonian of the system in which Rabi oscillations are induced, and $\hat{A}$ is assumed to be a Hermitian operator. 
Furthermore, $\Lambda$ represents the amplitude of the oscillating magnetic field, and $\omega$ denotes its frequency.
The state $\ket{E_0}$ represents the ground state, while $\ket{E_n}$ denotes the $n$-th excited state. The quantities $E_0$ and $E_n$ are the corresponding ground-state and excited-state energies, respectively.
\textcolor{black}{
Throughout this paper, we set ($\hbar=1$). Accordingly, all energy scales appearing in the Hamiltonian are expressed in angular-frequency units.
}
We define $N=2^L$, where $L$ is the number of qubits. 
We define the Pauli operators in the two-level subspace as
\begin{align}
\hat{Z} &= \ket{E_1}\bra{E_1} - \ket{E_0}\bra{E_0},\\
\hat{X} &= \ket{E_1}\bra{E_0} + \ket{E_0}\bra{E_1}.
\end{align}
The effective Hamiltonian $\hat{H}_{\rm eff}$ is defined as
\begin{align}
\hat{H}_{\rm eff}
=
\hat{U}\hat{H}\hat{U}^{\dagger}
+
i\frac{\partial \hat{U}}{\partial t}\hat{U}^{\dagger},
\end{align}
where $\hat{U}=e^{i\hat{H}_0 t}$. We assume that the energy difference between the ground state $\ket{E_0}$ and the first excited state $\ket{E_1}$ satisfies
\begin{align}
E_1-E_0 \simeq \omega,
\end{align}
and that all other energy differences satisfy
\begin{align}
|E_n-E_m-\omega| \gg \Lambda
\left|
\bra{E_n}\hat{A}\ket{E_m}
\right|.
\end{align}
Under these conditions, the effective Hamiltonian can be approximated as
\begin{align}
\hat{H}_{\rm eff}
&\approx
\frac{1}{2}\Lambda
e^{i(E_1-E_0-\omega)t}
\ket{E_1}\bra{E_1}
\hat{A}
\ket{E_0}\bra{E_0}
\notag\\
&\quad+
\frac{1}{2}\Lambda
e^{i(E_0-E_1+\omega)t}
\ket{E_0}\bra{E_0}
\hat{A}
\ket{E_1}\bra{E_1}.
\end{align}


Next, we define
\begin{equation}
\epsilon = E_1 - E_0 - \omega .
\end{equation}
We further introduce the unitary operator
\textcolor{black}{
\begin{equation}
\hat{U}'
=
e^{i\frac{\epsilon}{2}\hat{Z}t}.
\end{equation}
}
By transforming into the rotating frame using $\hat{U}'$, the Hamiltonian in the rotating frame, denoted by $\hat{H}'_{\rm eff}$, is given by
\begin{align}
\hat{H}_{\rm eff}'
&=
\hat{U}'\hat{H}_{\rm eff}\hat{U}'^{\dagger}
+
i\frac{\partial \hat{U}'}{\partial t}\hat{U}'^{\dagger}
\notag\\
&=
\frac{\epsilon}{2}\hat{Z}
+
\frac{\Lambda}{2}
|\bra{E_1}\hat{A}\ket{E_0}|
\hat{X}.
\end{align}
We then consider the following Hamiltonian:
\begin{align}
\hat{H}
=
\frac{\epsilon}{2}\hat{Z}
+
\frac{\Lambda'}{2}\hat{X}.
\end{align}
Here, $\hat{H}$ denotes the Hamiltonian of the system, $\epsilon$ represents the energy detuning, and $\Lambda'$ characterizes the effective strength of the external driving field, defined as $\Lambda'=\Lambda|\bra{E_1}\hat{A}\ket{E_0}|$.

Next, we define the parameters $r$ and $\theta$ as
\begin{align}
\Lambda' &= r\sin{2\theta},\\
\epsilon &= r\cos{2\theta}.
\end{align}
Here, $r$ characterizes the overall strength of the Hamiltonian, while $\theta$ determines the relative magnitudes of $\epsilon$ and $\Lambda'$.These definitions provide an explicit representation of how the external driving field affects the energy difference of the system.
We now consider the time-evolved states
\begin{align}
\ket{\psi_1}=e^{-i\hat{H}t}
\ket{E_1},
\end{align}
and
\begin{align}
\ket{\psi_0}=e^{-i\hat{H}t}
\ket{E_0},
\end{align}
which originate from the initial states $\ket{E_1}$ and $\ket{E_0}$, respectively.
The expectation values of $\hat{Z}$ are then given by
\begin{align}
 \bra{\psi_1} \hat{Z} \ket{\psi_1} =
 \frac{
\epsilon^2
+
\Lambda'^2
\cos
\sqrt{\epsilon^2+\Lambda'^2}t
}{
\epsilon^2+\Lambda'^2
},
\end{align}
and
\begin{align}
 \bra{\psi_0} \hat{Z} \ket{\psi_0} =
-\frac{
\epsilon^2
+
\Lambda'^2
\cos
\sqrt{\epsilon^2+\Lambda'^2}t
}{
\epsilon^2+\Lambda'^2
}.
\label{kitaiti}
\end{align}
These expressions show that the expectation values oscillate in time. The oscillation frequency contains information about the energy gap $E_1-E_0$ between the ground state and the first excited state.

\section{D-Wave Quantum Annealer}

In this section, we describe the Hamiltonian implemented in the D-Wave quantum annealer. We first consider a one-dimensional chain for simplicity and later extend our analysis to a two-dimensional lattice.
We introduce a protocol for estimating the energy gap between the ground state and the first excited state of $\hat{H}(t)$ by using the annealing schedule provided by the D-Wave system. 
The parameters used to characterize the energy gap will be defined later.
The Hamiltonian of the transverse-field Ising model is given by

\begin{align}
\hat{H}(t) &= \frac{A(s(t))}{2}\hat{H}_{\rm d}
+\frac{B(s(t))}{2}\hat{H}_{\rm p},\\
\hat{H}_{\rm d} &= \sum_{i=1}^{L}\hat{\sigma}_{x}^{(i)},\\
\hat{H}_{\rm p} &= \sum_{i=1}^{L} g(t) h_i \hat{\sigma}_{z}^{(i)}
+\sum^{L-1}_{i=1}J_{i}\hat{\sigma}_{z}^{(i)}\hat{\sigma}_{z}^{(i+1)}.
\end{align}
Here, $\hat{\sigma}_{x,z}^{(i)}$ denote the Pauli operators acting on the $i$-th qubit, while $h_i$ and $J_{i}$ represent the local bias and coupling strength, respectively. In addition, $A(s)$ and $B(s)$ are functions of the annealing parameter $s$ (see Fig.~\ref{AB_s}), whereas $s$ and $g$ are functions of time $t$. Finally, $L$ denotes the number of qubits.

\section{Proposed Method}


In this section, we describe the proposed method for estimating the energy gap. In general, Rabi oscillations are induced by introducing a time-dependent term, $\cos \omega t$, into the Hamiltonian. However, since rapid magnetic-field modulation using microwaves is not available in the D-Wave system, it is not possible to directly apply a transverse magnetic field of the form $\cos \omega t$.

On the other hand, D-Wave allows the longitudinal magnetic field $g(t)$ to be controlled linearly as a function of time. In our proposal, we exploit this capability by periodically modulating $g(t)$ with a triangular waveform, thereby approximately realizing a periodically driven system. Using this method, which can be implemented on a D-Wave quantum annealer, we apply a magnetic-field modulation analogous to $\cos \omega t$ and attempt to reproduce the dynamical behavior corresponding to Rabi oscillations. Since a triangular wave contains a dominant Fourier component at the driving frequency, it can induce a resonance analogous to that of conventional Rabi oscillations.
\textcolor{black}{For clarity, we first define the computational basis states used throughout this paper as
\begin{equation}
\hat{\sigma}_z\ket{0}=\ket{0},
\qquad
\hat{\sigma}_z\ket{1}=-\ket{1}.
\end{equation}
}In the following, we set $J_{i}=J=-1$ and $h_i=h<0$. For the parameter set considered in this paper, the ground state of $\hat{H}_{\rm p}$ is given by \textcolor{black}{$\ket{00\cdots0}$}, while the first excited state corresponds to \textcolor{black}{$\ket{11\cdots1}$}. The system is then evolved under $\hat{H}(t)$.

\begin{figure}[H] 
\centering 
\includegraphics[width=0.9\linewidth]{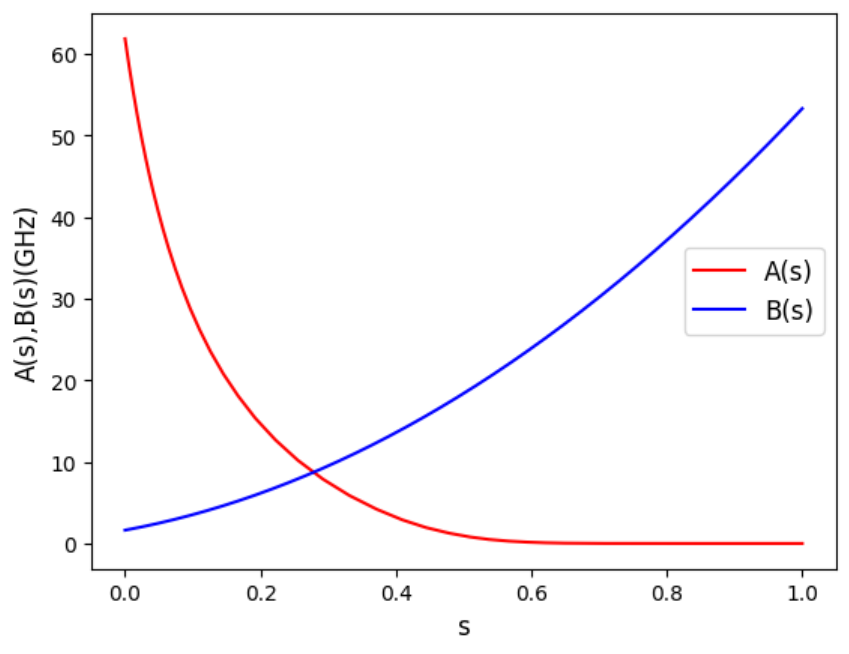}
\caption{
Annealing schedules $A(s)$ and $B(s)$ as functions of $s$ for the standard annealing schedule of the D-Wave Advantage\_system4.1 solver~\cite{DWaveSchedule2026}.
The horizontal axis represents $s$, while the vertical axis represents $A(s)$ (GHz) and $B(s)$ (GHz).
}
\label{AB_s} 
\end{figure}
\begin{figure}[H] 
\centering 
\includegraphics[width=0.9\linewidth]{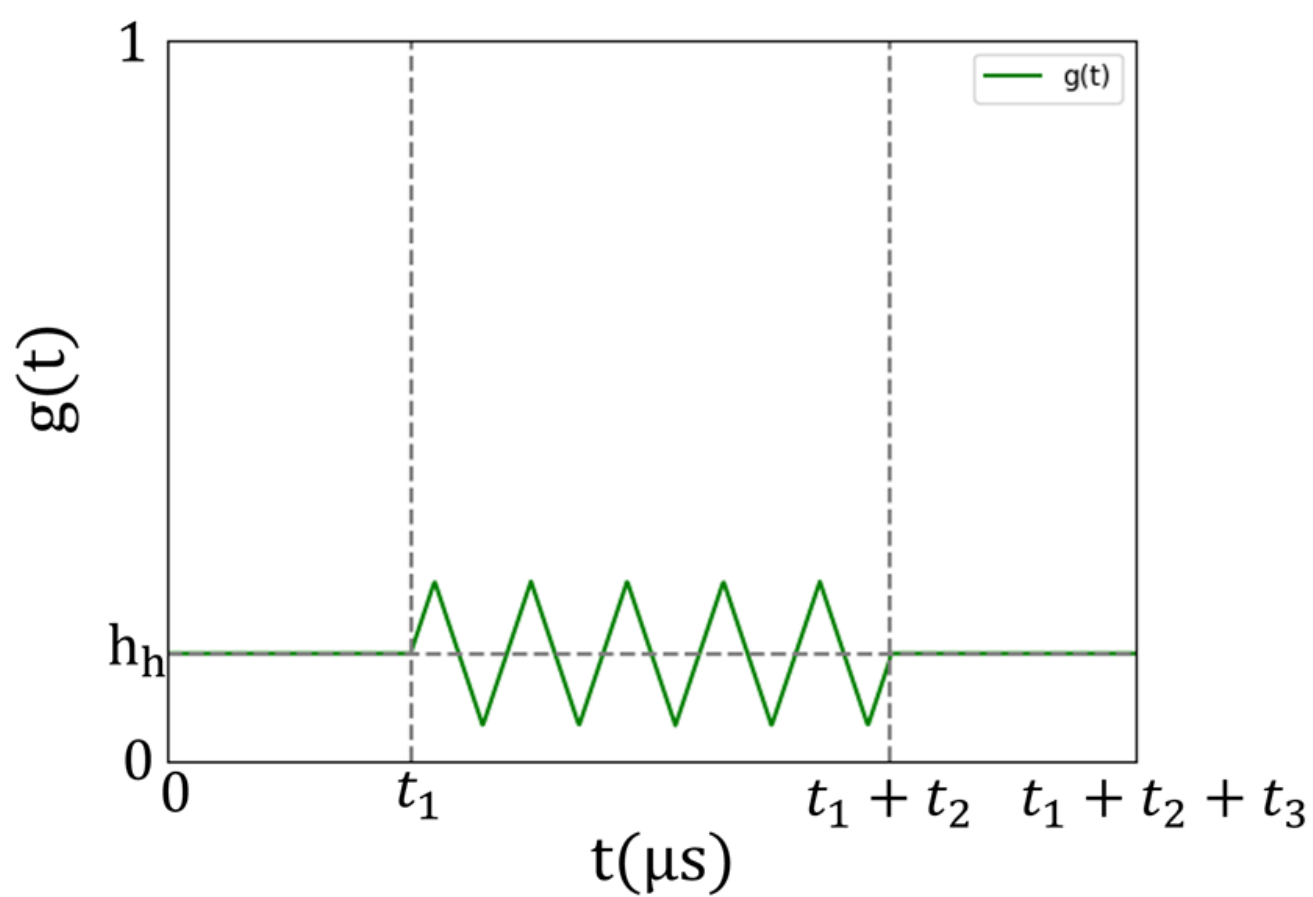}
\caption{Value of $g(t)$ as a function of time $t$.
$h_h$ denotes the central value of the oscillating magnetic field.}
\label{gsche2} 
\end{figure}
\begin{figure}[H] 
\centering 
\includegraphics[width=0.9\linewidth]{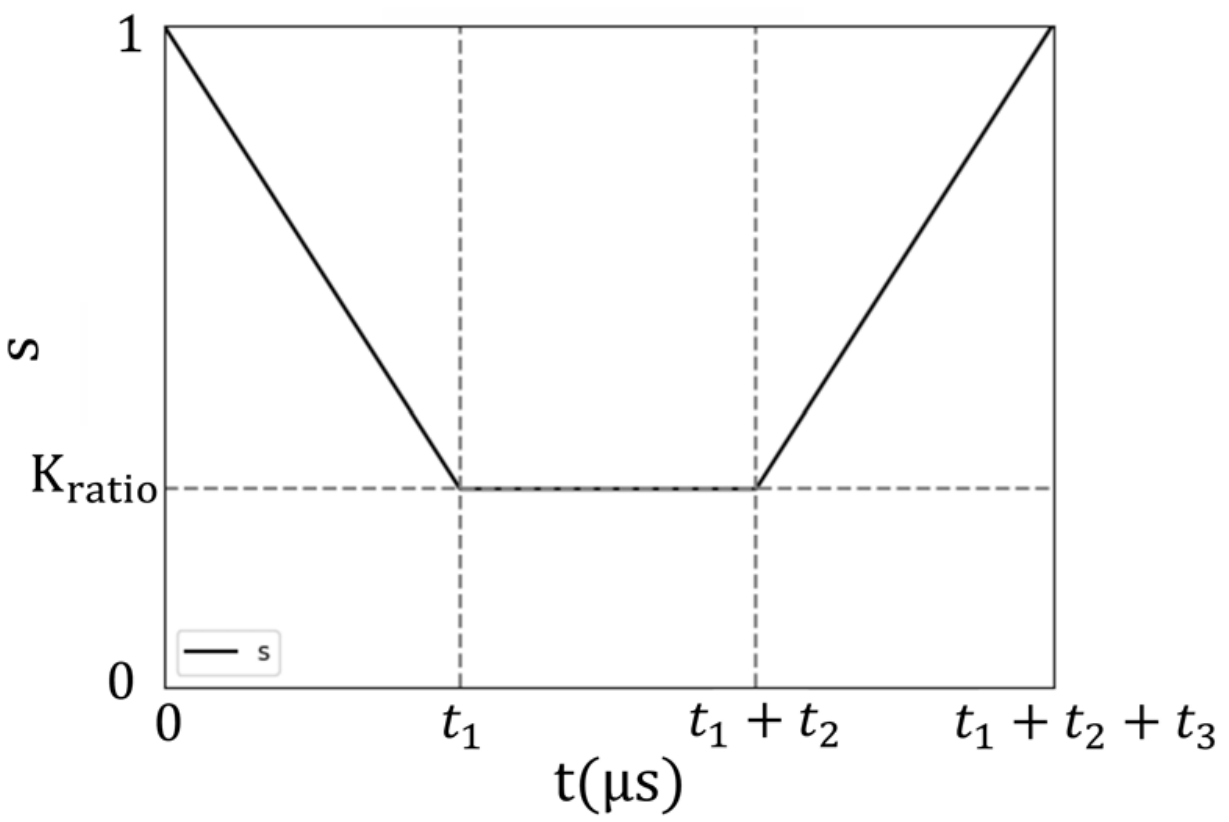}
\caption{Value of $s$ as a function of time $t$.
$K_{\rm{ratio}}$ denotes the minimum value of $s$ during the annealing schedule.}
\label{ts} 
\end{figure}

The schedule of $g(t)$ as a function of time is shown in Fig.~\ref{gsche2}, the schedule of $s$ as a function of time is shown in Fig.~\ref{ts}, and the dependence of $A(s)$ and $B(s)$ on $s$ is shown in Fig.~\ref{AB_s}. The schedule of $g(t)$ is given by Eq.~\eqref{gt}.
{\scriptsize
\begin{equation}
\label{gt}
g(t) =
\begin{cases}
h_h, & 0 \le t < t_1, \\[3pt]
h_h + \dfrac{2\lambda}{\pi}
    \arcsin\!\big[\sin\!\big(\omega_{\mathrm{drive}} (t - t_1)\big)\big],
    & t_1 \le t < t_1 + t_2, \\[3pt]
h_h, & t_1 + t_2 \le t < t_1 + t_2 + t_3.
\end{cases}
\end{equation}}
Here, $\lambda$ denotes the amplitude of the triangular wave, $h_{\rm{h}}$ is the center value of the oscillating magnetic field, and $\omega_{\rm{drive}}$ is the frequency of the triangular wave, defined as \textcolor{black}{$\omega_{\rm{drive}}=\frac{2\pi n_{\rm{cyc}}}{t_2}$, where $n_{\rm{cyc}}$} represents the number of oscillation cycles.
In the first and third intervals of Eq.~\eqref{gt}, $g(t)$ remains fixed at the constant value $h_h$. In the second interval, a periodic triangular-wave component is applied. This waveform is constructed using the $\sin$ and $\arcsin$ functions, resulting in a signal that repeatedly increases and decreases linearly. Consequently, the interval $t_1 \le t < t_1+t_2$ contains a schedule with $n_{\rm{cyc}}$ cycles of linear oscillation. This schedule enables modulation of the parameter $g(t)$ during the second interval.

It should be noted that the parameter $\lambda$ in Eq.~\eqref{gt} does not directly correspond to the amplitude of the resonant driving component. Expanding the triangular-wave term into a Fourier series yields
\begin{equation}
\frac{2\lambda}{\pi}
\arcsin\!\big[\sin(\omega_{\rm drive} t)\big]
=
\frac{8\lambda}{\pi^2}
\sum_{m=0}^{\infty}
\frac{(-1)^{m}\sin[(2m+1)\omega_{\rm drive}t]}
     {(2m+1)^2}.
\end{equation}
Therefore, the amplitude of the fundamental frequency component is given by $8\lambda/\pi^2$. Since the Rabi oscillation is predominantly driven by this resonant component, the effective coupling strength is
\begin{equation}
\lambda_{\rm eff}
=
\frac{8\lambda}{\pi^2}
\left|
\bra{E_1}\hat{A}\ket{E_0}
\right|.
\end{equation}
Accordingly, the Rabi frequency is determined by $\lambda_{\rm eff}$ rather than by $\lambda$ itself.
For the purpose of interpreting the resonance spectra, we assume that resonances induced by the higher harmonics of the triangular wave are sufficiently detuned and make negligible contributions near the fundamental resonance. Nevertheless, all Fourier components of the triangular wave are retained in the numerical simulations.
The current D-Wave device has a time resolution of approximately 10 ns, and therefore a modulation timescale of about 10 ns is feasible with existing hardware technology. The annealing parameter $s(t)$ is given by Eq.~\eqref{st}.
{\footnotesize
\begin{equation}
\label{st}
s(t)=
\begin{cases}
1-\dfrac{1-K_{\rm ratio}}{t_1}t,
& 0 \le t < t_1, \\[4pt]
K_{\rm ratio},
& t_1 \le t < t_1+t_2, \\[4pt]
K_{\rm ratio}
+\dfrac{1-K_{\rm ratio}}{t_3}
\left(t-t_1-t_2\right),
& t_1+t_2 \le t < t_1+t_2+t_3 .
\end{cases}
\end{equation}
}
where $K_{\rm ratio}$ denotes the minimum value of the annealing parameter during the schedule.

Finally, since measurements cannot be performed while the transverse-field term is present in the D-Wave system, the values of $A(s)$ and $B(s)$ are returned from their values at $t=t_1+t_2$ to those corresponding to the Ising Hamiltonian at $t=0$ during the interval from $t=t_1+t_2$ to $t=t_1+t_2+t_3$, after which the measurement is performed.



\section{Numerical Results}

\begin{figure}[H]
\centering
\includegraphics[width=0.95\linewidth]{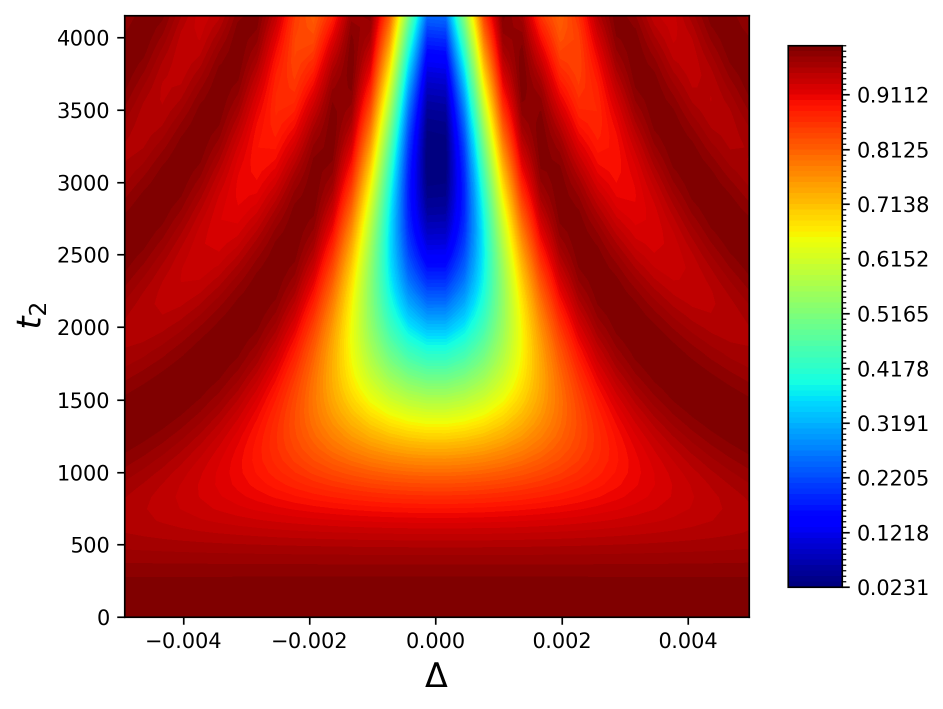}
\caption{Occupation probability of the ground state at $t=t_1+t_2+t_3$.
The $x$-axis represents $\Delta=\omega_{\rm{drive}}-\omega_{01}$, the $y$-axis represents the duration of the applied oscillating magnetic field $t_2$, and the $z$-axis represents the occupation probability of the ground state. Here, $\omega_{\rm{drive}}$ denotes the driving frequency of the triangular wave, and $\omega_{01}=0.756580(\mathrm{GHz})$ is the energy difference between the ground state and the first excited state.
The parameters are set to
$L=2,\lambda=0.01,J=-1,h=-0.01,h_h=1,K_{\rm{ratio}}=0.4004004,t_{1}=t_{3}=240(\mathrm{ns})$.
}
\label{Ground}
\end{figure}

\begin{figure}[H]
\centering
\includegraphics[width=0.95\linewidth]{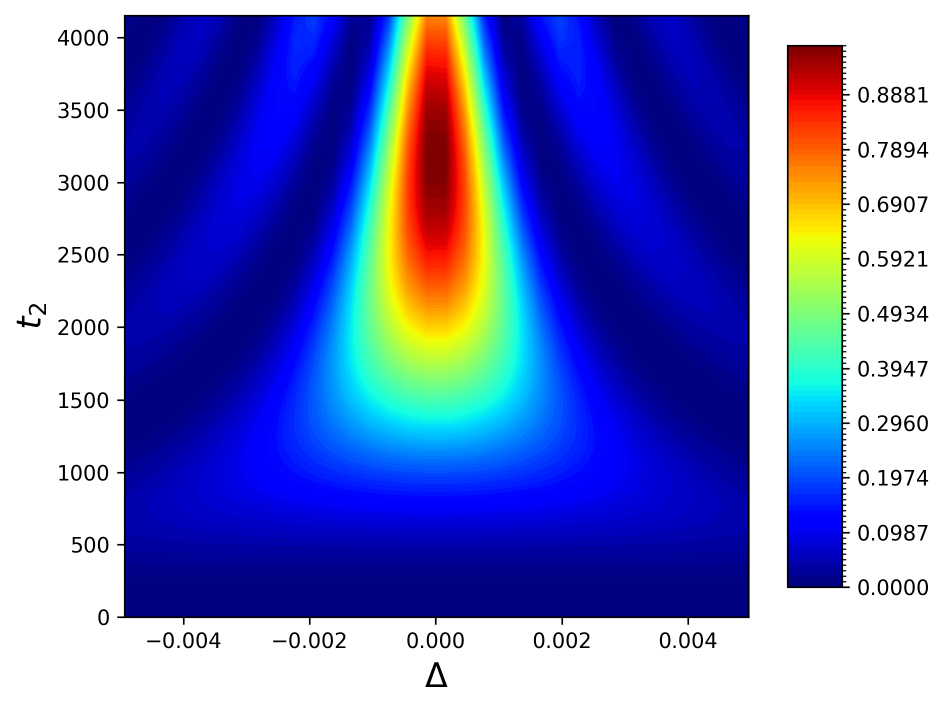}
\caption{Occupation probability of the first excited state at $t=t_1+t_2+t_3$.
The $x$-axis represents $\Delta=\omega_{\rm{drive}}-\omega_{01}$, the $y$-axis represents the duration of the applied oscillating magnetic field $t_2$, and the $z$-axis represents the occupation probability of the first excited state. Here, $\omega_{\rm{drive}}$ denotes the driving frequency of the triangular wave, and $\omega_{01}=0.756580$(GHz) represents the energy difference between the ground state and the first excited state.
The parameters are the same as those used in Fig.~\ref{Ground}.}
\label{First}
\end{figure}

We evaluate the performance of the proposed method described in the previous section through numerical simulations. 
\textcolor{black}{The initial state is the ground state of $\hat{H}_{\rm p}$ at $t=0$.}
By varying the driving frequency $\omega_{\mathrm{drive}}$ and the duration of the applied triangular-wave field $t_2$, we measure the populations of the ground state and the first excited state of $H_{\rm{p}}$ at the end of the protocol. The results for the two-qubit case are shown in Figs.~\ref{Ground} and \ref{First}.

From Figs.~\ref{Ground} and \ref{First}, it can be seen that the ground-state population decreases to nearly zero around the detuning $\Delta=0$, while the population of the first excited state simultaneously increases to nearly unity. This result indicates that Rabi resonance occurs when the driving frequency $\omega_{\mathrm{drive}}$ matches the intrinsic transition frequency $\omega_{01}$ of the system, resulting in the most efficient transition from the ground state to the first excited state. In other words, a nearly complete population transfer from the ground state to the first excited state occurs at $\Delta=0$, providing a clear signature of the resonance condition.

To understand these numerical results, we reconsider the time dependence of the expectation value $\bra{\psi_0}\hat{Z}\ket{\psi_0}$ derived in the previous section [Eq.~\eqref{kitaiti}]. This expression implies that the system exhibits oscillatory behavior determined by $\epsilon$ and \textcolor{black}{$\Lambda'$ during} time evolution. In particular, when the angular frequency of the applied driving field $\omega_{\mathrm{drive}}$ coincides with the transition frequency $\omega_{01}$ of the system, namely under the resonance condition
\begin{equation}
\Delta = \omega_{\mathrm{drive}} - \omega_{01} = 0,
\end{equation}
the transition probability reaches its maximum.

To quantitatively characterize this resonance behavior, we obtained the population of the ground state,
$P(\omega_{\rm drive})$, while sweeping the driving frequency $\omega_{\rm drive}$.
The resulting spectral response was then fitted using a Lorentzian function in order to determine the resonance frequency.
Specifically, we employed the following fitting function.

\begin{align}
    P(\omega_{\mathrm{drive}}) 
    = P_{\mathrm{max}} 
      - (P_{\mathrm{max}} - P_{\mathrm{min}}) 
      \frac{\lambda_{\rm{fit}}^2}
           {(\omega_{\mathrm{drive}} - \omega_{01}^{\rm{(fit)}})^2 + \lambda_{\rm{fit}}^2}.
      \label{lorentz_fit}
\end{align}
Here, $\omega_{01}^{\rm{(fit)}}$, $\lambda_{\rm{fit}}$, 
$P_{\mathrm{max}}$, and $P_{\mathrm{min}}$
are fitting parameters. The parameter $\omega_{01}^{\rm{(fit)}}$ represents the estimated energy gap between the ground state and the first excited state, while $\lambda_{\rm{fit}}$ represents the estimated linewidth.
The fitting results obtained are as follows:
\begin{center}
\begin{tabular}{ll}
$\omega_{01}^{\rm{(fit)}}$ & $= 0.756586$(GHz) 
\end{tabular}
\end{center}
From these results, it can be seen that the transition probability is maximized, or equivalently the ground-state population is
minimized, around $\omega_{\mathrm{drive}} \simeq 0.756586$. Furthermore, since the exact value of $\omega_{01}$ is 0.756580, the proposed method is capable of estimating the energy gap with high accuracy.

\begin{figure}[H]
\centering
\includegraphics[width=0.95\linewidth]{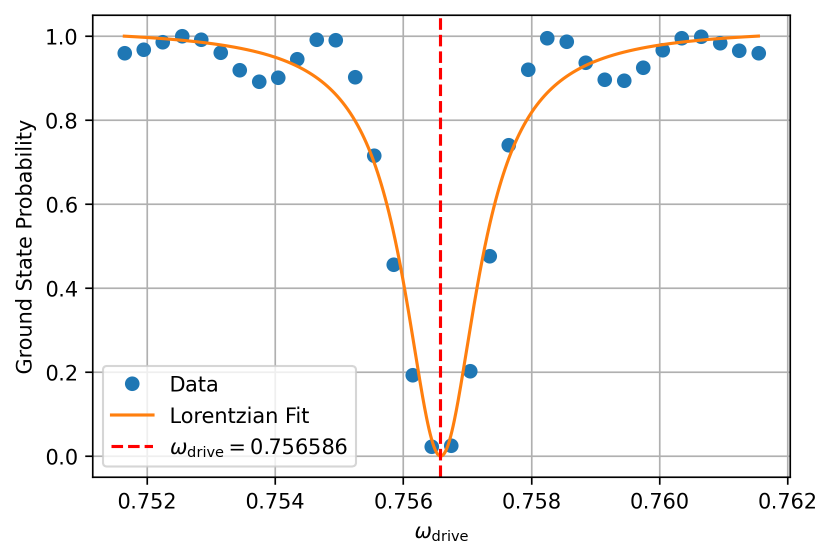}
\caption{
Results of the Lorentzian fitting.
The $x$-axis represents the driving frequency of the oscillating magnetic field, $\omega_{\rm{drive}}$, and the $y$-axis represents the occupation probability of the ground state. The parameter $t_2$ is set to $3155.792(\mathrm{ns})$, while all other parameters are identical to those used in Fig.~\ref{Ground}.
}
\label{lorentzian}
\end{figure}

Figure~\ref{lorentzian} shows the measured data (points) together with the fitting curve (solid line) obtained using Eq.~\eqref{lorentz_fit}.
As described above, the fitting results reproduce well our numerical plots,
confirming the characteristic behavior of Rabi oscillations, namely that the transition probability becomes maximal in the vicinity of the resonance condition $\omega_{\mathrm{drive}}=\omega_{01}$.



\begin{figure}[H]
\centering
\includegraphics[width=0.95\linewidth]{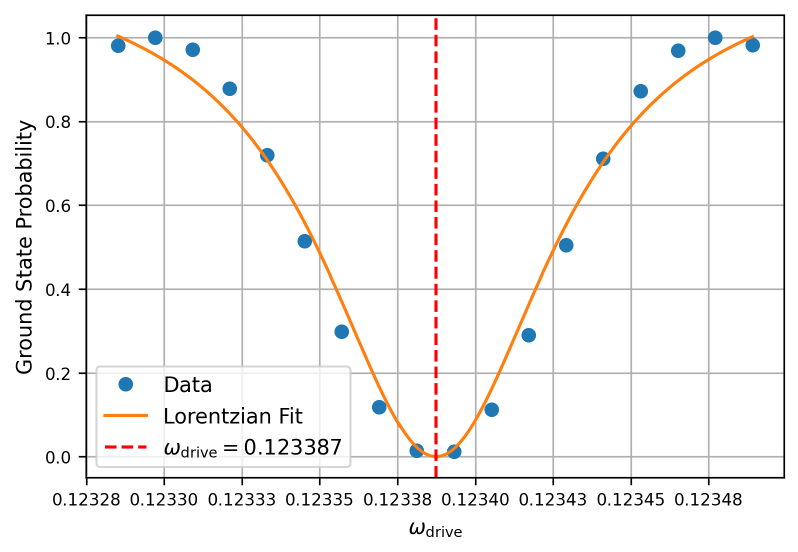}
\caption{
Results of the Lorentzian fitting.
The $x$-axis represents the driving frequency of the oscillating magnetic field, $\omega_{\rm{drive}}$, and the $y$-axis represents the occupation probability of the ground state. The parameter $t_2$ is set to $60394.437(\mathrm{ns})$, and $\omega_{01}=0.123386(\mathrm{GHz})$ denotes the energy gap between the ground state and the first excited state.
The remaining parameters are set to
$L=7,\lambda=0.00015,J=-1,h=-0.15,h_h=0.01,K_{\rm{ratio}}=0.3663664,t_{1}=t_{3}=6336(\mathrm{ns})$.
The fitted parameters are obtained as
$\omega_{01}^{\rm{(fit)}}=0.123387(\mathrm{GHz})$
and
$\lambda_{\rm{fit}}=0.000045$.
}
\label{q7_lorentzian}
\end{figure}

Also, we consider a seven-qubit system.
Figure~\ref{q7_lorentzian} shows the fitting result obtained for the seven-qubit system.
As shown in Fig.~\ref{q7_lorentzian}, the Lorentzian fitting remains in good agreement with the numerical data for the seven-qubit system, and the transition probability is maximized in the vicinity of the resonance frequency. 
Furthermore, for the parameter sets considered here, the seven-qubit system exhibits a smaller energy gap than the two-qubit system.

\subsection{Application to a Two-Dimensional Lattice System}

\begin{figure}[H]
\centering
\includegraphics[width=0.8\linewidth]{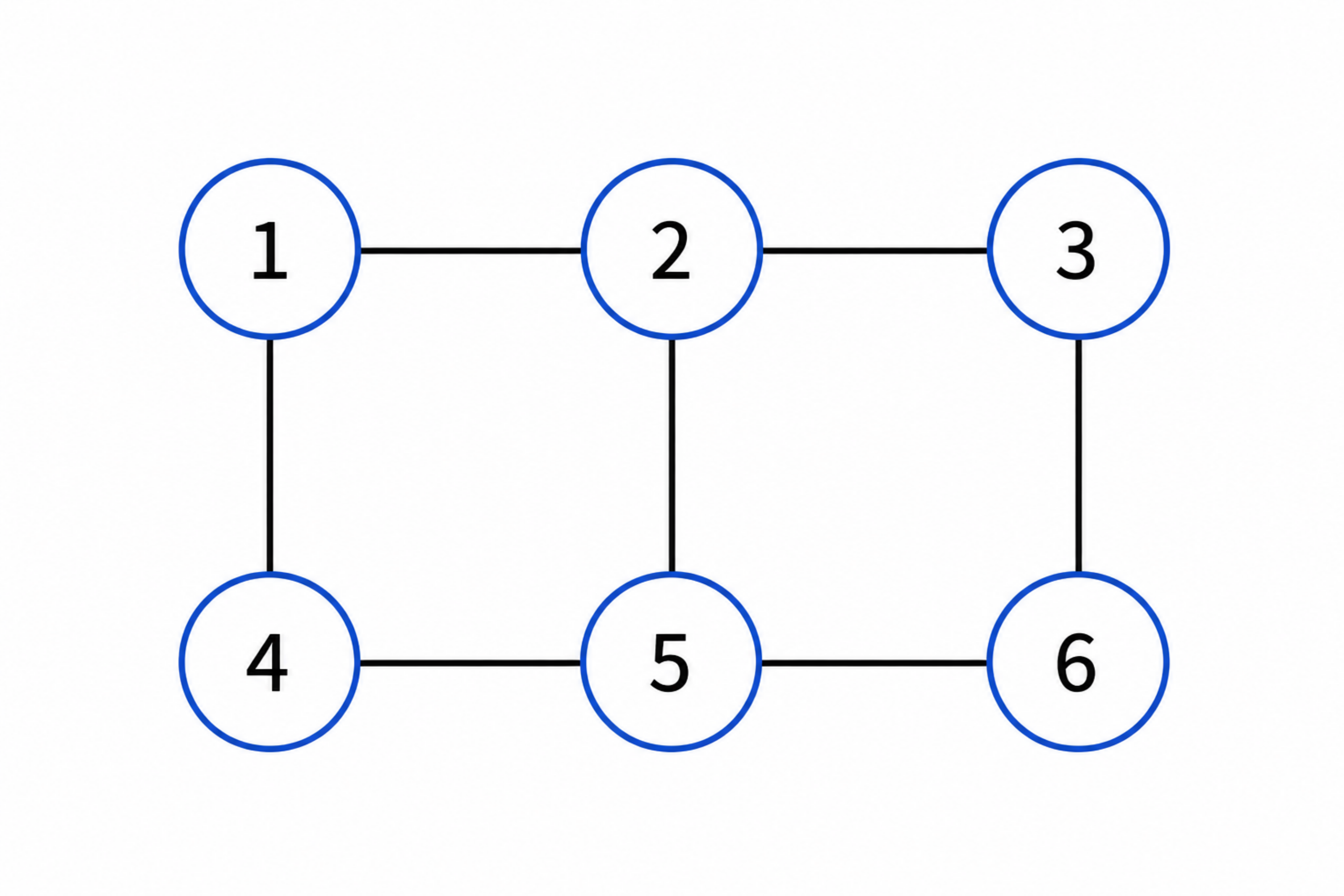}
\caption{
Interaction topology of the two-dimensional lattice system considered in this study. The six qubits are arranged on a $2\times3$ lattice, and nearest-neighbor interactions are introduced along the horizontal and vertical directions.
}
\label{q6_2dlattice}
\end{figure}

In addition to the \textcolor{black}{one-dimensional chain} considered thus far, we also analyzed a transverse-field Ising model with nearest-neighbor interactions on a two-dimensional lattice in order to examine the applicability of the proposed method to other interaction topologies. In this case, the problem Hamiltonian is given by

\begin{equation}
\hat{H}_{\rm p}=
\sum_{i=1}^{L}
g(t)h_i\hat{\sigma}_z^{(i)}
+J\sum_{\langle i,j\rangle}
\hat{\sigma}_z^{(i)}
\hat{\sigma}_z^{(j)}
\end{equation}
where $\langle i,j\rangle$ denotes a pair of nearest-neighbor sites on the two-dimensional lattice.
\textcolor{black}{
Figure~\ref{q6_2dlattice} illustrates the interaction topology considered in this study. We employ a $2\times3$ lattice consisting of six qubits, where each qubit interacts only with its nearest neighbors.
}

\begin{figure}[H]
\centering
\includegraphics[width=0.95\linewidth]{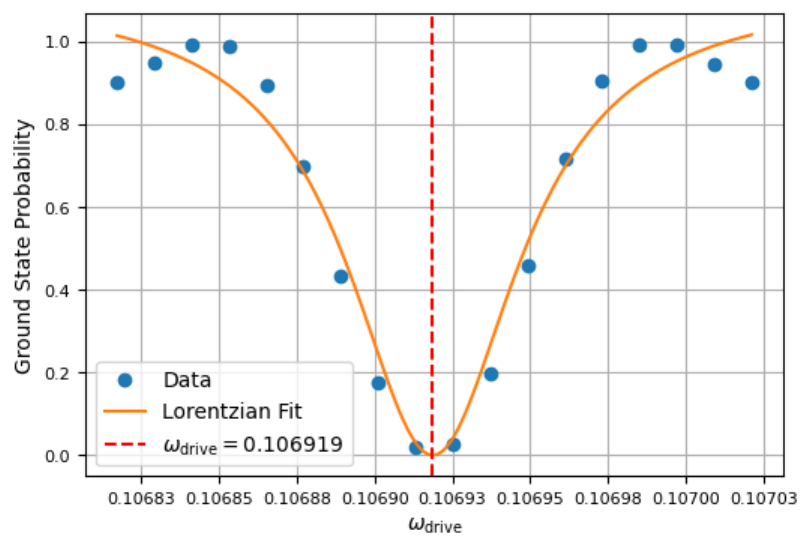}
\caption{
Results of the Lorentzian fitting.
Here, a $2\times3$ two-dimensional lattice is considered.
The $x$-axis represents the driving frequency of the oscillating magnetic field, $\omega_{\rm{drive}}$, and the $y$-axis represents the occupation probability of the ground state. The parameter $t_2$ is set to $77101.103(\mathrm{ns})$, and $\omega_{01}=0.106918(\mathrm{GHz})$ denotes the energy gap between the ground state and the first excited state.
The remaining parameters are set to
$L=6,\lambda=0.00002,J=-1,h=-0.15,h_h=0.01,K_{\rm{ratio}}=0.3183183,t_{1}=t_{3}=6817(\mathrm{ns})$.
The fitted parameters are obtained as
$\omega_{01}^{\rm{(fit)}}=0.106919(\mathrm{GHz})$
and
$\lambda_{\rm{fit}}=0.000033$.
}
\label{q6_2dlorentzian}
\end{figure}

Figure~\ref{q6_2dlorentzian} shows the fitting result obtained for the two-dimensional lattice system.
As shown in Fig.~\ref{q6_2dlorentzian}, the transition probability increases in the vicinity of the resonance frequency. Furthermore, the Lorentzian fitting is in good agreement with the numerical data, indicating that the energy gap can be estimated from the resonance frequency.
These results demonstrate that the proposed method is also applicable to systems with nearest-neighbor interactions on a two-dimensional lattice. This suggests that the method is effective for quantum spin systems with different interaction topologies.

\textcolor{black}{
Next, we discuss the limitations associated with applying the proposed method to an actual D-Wave device. Since the present approach estimates the energy gap based on the resonance condition $\omega_{\mathrm{drive}} \simeq \omega_{01}$, the transition frequency $\omega_{01}$ of the system must lie within the range of driving frequencies that can be realized on the hardware. Owing to the limitations of the D-Wave device, oscillations with periods as short as $\pi/\omega_{\rm drive} \sim 10(\mathrm{ns})$ can be implemented, corresponding to an angular frequency of approximately $\omega \lesssim 0.314(\mathrm{GHz})$.} 

\begin{figure}[H]
\centering
\includegraphics[width=0.90\linewidth]{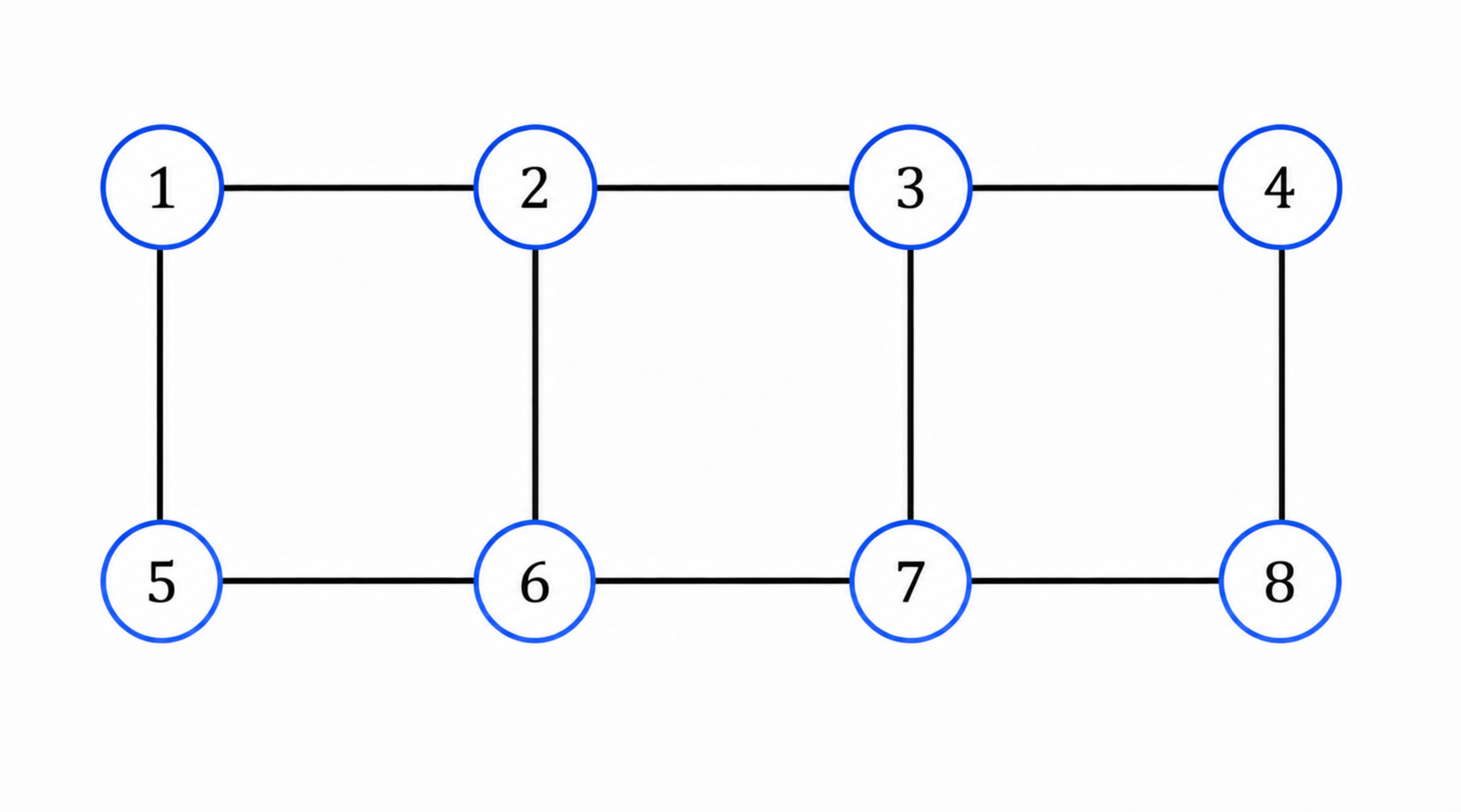}
\caption{
Interaction topology of the eight-qubit two-dimensional lattice considered in this study. The qubits are arranged on a $2\times4$ lattice, and nearest-neighbor interactions are introduced along the horizontal and vertical directions.
}
\label{q8_2dlattice}
\end{figure}

\textcolor{black}{
Moreover, this minimum programmable timescale does not directly determine the angular-frequency range over which the desired waveform can be faithfully delivered to the qubits. In the D-wave system, the linear-bias-control waveforms delivered to the qubits pass through a low-pass filter with a cutoff angular frequency of around $40$ MHz~\cite{DWaveAnnealingControls}. This means that a triangular-wave oscillating magnetic field with a small period
could be distorted.}

Therefore, for a faithful implementation of the proposed method, the transition angular frequency $\omega_{01}$ should preferably lie well below $40$ MHz. In our previous numerical simulations, however, the energy gaps were too large to satisfy this requirement. To address this issue, we investigate parameter regimes in which $\omega_{01}$ lies well below $40$ MHz and is therefore expected to be experimentally accessible.

\begin{figure}[H]
\centering
\includegraphics[width=0.95\linewidth]{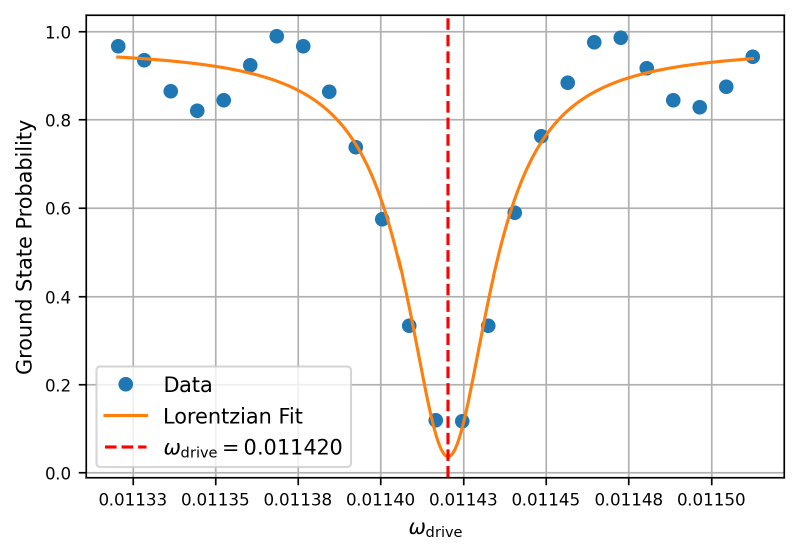}
\caption{
Results of the Lorentzian fitting.
Here, a $2\times4$ two-dimensional lattice is considered.
The $x$-axis represents the driving frequency of the oscillating magnetic field, $\omega_{\rm drive}$, and the $y$-axis represents the occupation probability of the ground state.
The parameter $t_2$ is set to $135896.583(\mathrm{ns})$, and $\omega_{01}=0.011420(\mathrm{GHz})$ denotes the energy gap between the ground state and the first excited state.
The remaining parameters are set to
$L=8,\lambda=0.00002,J=-1,h=-0.05,h_h=0.002,K_{\rm ratio}=0.3183183,t_{1}=t_{3}=13634(\mathrm{ns})$.
The fitted parameters are obtained as
$\omega_{01}^{\rm(fit)}=0.011420(\mathrm{GHz})$
and
$\lambda_{\rm fit}=$0.000016.
}
\label{q8_2dlorentzian}
\end{figure}

\textcolor{black}{
To investigate this parameter regime, we extend the system considered above from a six-qubit ($2\times3$) lattice to an eight-qubit ($2\times4$) lattice, while employing the same transverse-field Ising Hamiltonian despite the difference in lattice size.
Figure~\ref{q8_2dlattice} illustrates the interaction topology considered in this analysis. As in the six-qubit case, only nearest-neighbor interactions are introduced along the horizontal and vertical directions.
\textcolor{black}{
Figure~\ref{q8_2dlorentzian} shows the fitting result obtained for the eight-qubit system.
As in the numerical calculations presented above, the Lorentzian fit accurately reproduces the resonance profile, demonstrating that the energy gap can be estimated from the resonance frequency.
}}

\textcolor{black}{
In the present numerical simulations, the modulation function ($g(t)$) is composed of 250 periods of a triangular waveform. On the D-Wave Advantage system 4.1 solver, however, ($g(t)$) must be implemented using the 
\texttt{h\_gain\_schedule} parameter, which specifies the time-dependent gain applied to the linear bias terms~\cite{DWaveSolverParameters}. 
  For the solver employed in this study, the maximum number of programmable schedule points, specified by the solver property \texttt{max\_h\_gain\_schedule\_points}, is limited to 20~\cite{DWaveSolverProperties}. 
 This constraint restricts the presently implementable waveform to approximately eight periods of the triangular modulation. Nevertheless, this 
limitation originates from the current hardware and control interface rather than from the proposed method itself. Future improvements in waveform programmability, such as an increased number of available schedule points, could enable the implementation of a substantially larger number of modulation periods.
}


\section*{Conclusion}


In conclusion, we propose a method for estimating the energy gap between the ground state and the first excited state in the transverse-field Ising model by utilizing Rabi oscillations induced by a triangular-wave modulation. By sweeping the driving frequency and analyzing the population dynamics under resonant conditions, we show that the intrinsic transition frequency of the system can be effectively extracted. A key feature of the proposed method is that it does not rely on eigenvalue calculations and instead directly probes the excitation structure through the dynamical response of the system. Furthermore, the method is applicable to large-scale quantum spin systems, provided that the relevant transition frequency lies within the experimentally accessible modulation bandwidth and that the ramp-back process preserves the population information. Under these conditions, it provides a promising alternative approach for overcoming the computational limitations of classical numerical methods.
In the present study, decoherence was neglected in order to clarify the fundamental operating principle of the proposed method. Future work will investigate the influence of realistic decoherence. Extending the present analysis to larger multi-qubit systems and nonadiabatic regimes will be important for assessing the generality and practical applicability of the method.

This project is supported by
JST Moonshot R\&D Grant
Number JPMJMS226C, 
JST CREST Grant Number JPMJCR23I5, and Presto
JST Grant Number JPMJPR245B.
The authors would like to thank Y. Suzuki and T. Kadowaki for helpful discussions.

\bibliography{ref}

\end{document}